\documentclass[11pt]{article}
\usepackage{moriond,epsfig}

\def\Journal#1#2#3#4{{#1} {\bf #2}, #3 (#4)}

\def\NPB{{\em Nucl. Phys.} B}

\def\APP{\em Astroparticle Physics}

\def\cp{\mbox{$\delta_{CP}$\,}}
\def\tai{\mbox{$\theta_{13}$\,}}
\def\ssq{\mbox{$\sin^{2}(2 \theta_{13})$\,}}
\def\t12a{\mbox{$\theta_{12}$}}
\def\t23a{\mbox{$\theta_{23}$}}
\def\tsol{\mbox{$\theta_{sol}$}}
\def\tatm{\mbox{$\theta_{atm}$}}
\def\dmsol{\mbox{$\Delta m^{2}_{sol}$}}
\def\dmatm{\mbox{$\Delta m^{2}_{atm}$}}
\def\dms{\mbox{$\Delta m^{2}_{12}$}}
\def\dmss{\mbox{$\Delta m^{2}_{23}$}}
\def\dmsss{\mbox{$\Delta m^{2}_{13}$}}
\def\dc{{\em Double-Chooz }}

\def\be{\begin{equation}}
\def\ee{\end{equation}}
\def\bea{\begin{eqnarray}}
\def\eea{\end{eqnarray}}

\begin{document}
\vspace*{4cm}
\title{Theta 13 Determination with Nuclear Reactors}

\author{ F. Dalnoki-Veress for the Double-Chooz Project}

\address{Max-Planck-Institut f\"{u}r Kernphysik, Saupfercheckweg 1 \\
D-69117, Heidelberg, Germany}

\maketitle\abstracts{Recently there has been a lot of interest
around the world in the use of nuclear reactors to measure \tai,
the last undetermined angle in the 3-neutrino mixing scenario. In
this paper the motivations for \tai measurement using short
baseline nuclear reactor experiments are discussed. The features
of such an experiment are described in the context of {\it Double
Chooz}, which is a new project planned to start data-taking in
2008, and to reach a sensitivity of \ssq$<$0.03.}

\section{Current State of Neutrino Physics}
The past few decades have seen much progress in the field of
neutrino physics. We now believe that active neutrinos mix in
three flavours as quarks do in the quark sector, and that as a
consequence the neutrinos must have mass. However, neutrinos mix
with very different mixing angles from the quark sector since we
know two of the mixing angles are large, whereas the mixing angle
in the quark sector is exceedingly small. In the three neutrino
mixing~\footnote{The assumption in this note is that only 3
neutrino types exist. If MiniBoone confirms the existence of a
fourth neutrino the current scenario will need to be re-evaluated}
scenario the mixing between the mass ($\nu_{i}$ where i=1,2,3) and
flavour eigenstates ($\nu_{\alpha}$ where alpha represents the 3
active lepton flavours) can be parameterized in terms of a unitary
matrix often called the $U_{PMNS}$ mixing matrix ($\mathcal{M}$ is
the Majorana phase matrix). Nevertheless, the underlying symmetry
behind the structure of this matrix and its relation to the CKM
matrix is entirely unknown:
\begin{eqnarray}
U_{PMNS} &=& \underbrace{
\left(%
\begin{array}{ccc}
  1 &  &  \\
    & c23 & s23 \\
   &  -s23 & c23 \\
\end{array}%
\right) }_{\theta_{23}=\theta_{atmospheric}\sim 45^{\circ}}
\underbrace{
\left(%
\begin{array}{ccc}
  c13 &  & s13e^{-i\delta}  \\
   & 1 &  \\
  -s13e^{i\delta} &  & c13 \\
\end{array}%
\right)}_{\theta_{13}=\theta_{reactor} \leq 14.9^{\circ}}
\underbrace{
\left(%
\begin{array}{ccc}
  c12 & s12 &  \\
  -s12 & c12 &  \\
   &  & 1 \\
\end{array}%
\right)}_{\theta_{12}=\theta_{solar}\sim 28^{\circ}-39^{\circ}}
\mathcal{M}
\end{eqnarray}

Three types of experiments have been performed over the last 30
years: the pioneering solar neutrino measurements which first
identified problems in the neutrino sector, atmospheric neutrino
experiments, and short and long baseline nuclear reactor
experiments. These three types of experiments essentially measure
different components of the neutrino mixing matrix and the
factorized form of the neutrino mixing matrix can be used to
identify these different components. The three mixing angles are
labelled \tsol\, and \tatm\, according to the dominant type of
experiment that these angles are measured with. The third type of
experiment is the the short baseline neutrino experiments which
measure the final angle \tai\ in the 3 neutrino mixing scenario.
Using the mixing matrix the three flavour neutrino mixing
probability can be written as:
\begin{eqnarray}
P_{(\nu_{\alpha}\rightarrow\nu_{\beta})} & = &
\delta_{\alpha\beta}-2\Re \sum_{j > i} U_{\alpha i} U^{*}_{\alpha
j} U^{*}_{\beta i}U_{\beta j}\biggl(1-exp^{(i\Delta m^{2}_{ji} L /
2E)}\biggr)\label{eq:1} \\
& & where, \,\, \Delta m^{2}_{ji}=m_{j}^{2}-m_{i}^{2}
\end{eqnarray}
Therefore, an experiment measuring only one type of neutrino is
sensitive to all 3 neutrino mass eigenstates through the
$U_{PMNS}$ neutrino mixing matrix. The degree of sensitivity
depends on the type of experiment and the (L/E) distance between
the neutrino source and detector.

Neutrino oscillations are then described by three mixing angles
($\theta_{12}$, $\theta_{23}$, $\theta_{13}$), a CP violating
phase (\cp) and two quadratic mass splittings (\dms\, and \dmss).
Solar neutrino data combined with the results of the KamLAND long
baseline reactor experiment suggest $\dms=\dmsol\sim
6.9^{+2.6}_{-1.5} \times 10^{-5}$ , whereas atmospheric neutrino
detectors and K2K (a long baseline accelerator experiment) have
measured $\dmss=\dmatm=\pm 2^{+1.2}_{-0.9} \times 10^{-3}\sim
\dmsss$. An ambiguity exists in the sense that it is not known
whether m$_{1}$ (normal hierarchy) or m$_{3}$ (inverted hierarchy)
state is the lowest neutrino mass eigenstate. Future generations
of experiments especially double beta decay and long baseline
accelerator experiments will attempt to resolve this. Crucial to
this picture of neutrino mass mixing is the \tai\ mixing angle
which is not known and has only been constrained by reactor
experiments such as Chooz and Palo Verde to be $<$14.9$^{\circ}$
(3$\sigma$). For this reason it is imperative to measure
$\theta_{13}$ in the next generation of experiments.

\subsection{Measurement of Theta 13}
There are two types of experiments that will be sensitive to
$\theta_{13}$: the accelerator and nuclear reactor experiments.
The accelerator experiments produce pions and kaons which
primarily decay to produce $\nu_{\mu}$'s. These experiments are
typically long baseline experiments (LBL) optimized to the
distance where there is maximum conversion of $\nu_{\mu}$ to other
flavours. Using the formula above (\ref{eq:1}) and the
approximation that \dmatm=\dmsss, \tsol=$\theta_{12}$, and
\tatm=$\theta_{23}$ the conversion probability is
written~\cite{whitepaper}:
\begin{eqnarray}
P_{\nu_{\mu}\rightarrow \nu_{e}} &\simeq& \sin^{2}2\theta_{13}
\sin^{2}\theta_{atm} \sin^{2}(\Delta m^{2}_{atm}
L/ 4E) \\
 &\mp& \alpha \sin 2\theta_{13} \sin\delta_{CP} cos\theta_{13} \sin
 2\theta_{sol} \sin2\theta_{atm} (\Delta m^{2}_{atm} L/ 4E) \sin^{2} (\Delta m^{2}_{atm} L/ 4E) \label{eq:2}\\
 &+& \alpha \sin 2 \theta_{13} \cos \delta_{CP} \cos \theta_{13} \sin
 2\theta_{sol} \sin 2\theta_{atm} \cos (\Delta m^{2}_{atm} L/ 4E)
(\Delta m^{2}_{atm} L/ 4E) \nonumber \\
 && \sin(\Delta m^{2}_{atm} L/ 4E) \nonumber \\
 &+& \alpha^{2} \cos^{2} \theta_{atm} \sin^{2} 2\theta_{sol}(\Delta m^{2}_{atm} L/
 4E)^{2} \nonumber \\
 && where \; \alpha=\dmsol/\dmatm \simeq 0.029  \nonumber
\end{eqnarray}
The ($+$) sign in equation (\ref{eq:2}) refers to anti-neutrino
oscillations and the ($-$) sign refers to neutrino oscillations.
One notices that there are ambiguities in the sense that there are
parameter correlations and degeneracies due to different
combinations of parameters. For example, the survival probability
is the same if a substitution of $\theta_{23}\rightarrow
\bigl(\frac{\pi}{2}\bigr) - \theta_{23}$ is made, or one can vary
$\delta$ and $\theta_{13}$ to give the same overall survival
probability.

In nuclear reactor neutrino experiments the oscillation
probability is quite different~\footnote{Neglecting terms of
second order in $\alpha$}:
\begin{eqnarray}
P_{\overline{\nu_{e}}\rightarrow \overline{\nu_{e}}} &\simeq&
\underbrace{1-\sin^{2} 2\theta_{13} \sin^{2}(\Delta m^{2}_{atm}
L/4E)}_{\theta_{13}\, oscillations} + \alpha^{2} \underbrace{
(\Delta m^{2}_{atm} L/ 4E) \cos^{4}\theta_{13} \sin^{2}2
\theta_{sol}}_{\theta_{sol}\, influence}
\end{eqnarray}
Here only the first two terms contribute since $\alpha$ is small.
Furthermore, for small (L/E) the second term does not contribute
since the solar oscillation depletion occurs at much longer
baselines because of the small \dmsol. Therefore the survival
probability amounts to the first term of
$P_{\overline{\nu_{e}}\rightarrow \overline{\nu_{e}}}$, and
reduces to the familiar two neutrino oscillations survival
probability formula. Comparing the reactor neutrino survival
probability formula to the LBL probabilities, one notices that the
CP violating parameter \cp does not appear. This is the strength
of reactor experiments since these experiments allow $\theta_{13}$
to be measured independent of \cp without the ambiguities which
are present in LBL experiments. The CP asymmetry in LBL
experiments may be expressed as~\cite{jhfloi}:
    \begin{eqnarray}
    A_{CP}& = & \frac{P_{\nu_{\mu} \rightarrow \nu_{e}} -
    P_{\overline{\nu_{\mu}}\rightarrow
    \overline{\nu_{e}}}}
    {P_{\nu_{\mu} \rightarrow \nu_{e}} +
    P_{\overline{\nu_{\mu}}\rightarrow \overline{\nu_{e}}}} \nonumber \\
          & = & \biggl( \frac{\Delta m^{2}_{12} L \sin 2\theta_{12}}{4
    E_{\nu}} \biggr)\times \frac{\sin \delta_{CP} }{ \sin \theta_{13}}
    \end{eqnarray}
Consequently, determination of $\theta_{13}$ in nuclear reactor
experiments may also be useful for LBL experiments in tuning their
energy to maximize the \cp asymmetry.

\section{Theta 13 Reactor Experiments}
In the last few years there has been great interest in the
measurement of \tai using nuclear reactors. In fact an
international working group has been formed to study the
feasibility of such an experiment~\cite{whitepaper}. Currently,
there are several possible sites identified around the world for a
future reactor neutrino experiment. Table \ref{t:sites} shows a
summary of the possible sites and their specific characteristics.
\begin{table}[t]
\caption{Proposed sites around the world for a next generation
nuclear reactor neutrino project to measure $\theta_{13}$
~ \protect \cite{mg}.} \label{t:sites} \vspace{0.4cm}
\begin{center}
\begin{tabular}{|c|c|c|c|c|}
\hline
Site/Project & Reactor Power & Baseline & Overburden & Det. Volume \\
& (GW) & Near/Far (m) & Near/Far (m.w.e.) & Near/Far (t) \\
 \hline
 &  &  &  & \\
{\it Double-Chooz} (France) & 8.5 & 150/1050 & 60/300 &  10/10 \\
KR2DET (Russia) & 1.5 & 115/1000 & 600/600 & 45/45 \\
Diablo Canyon (USA) & 7 & 400/1800 & 100/700 & 25/50 \\
Angra (Brazil) & 4 & 350/1350 & 60/600 & 50/50 \\
Braidwood (USA) & 7 & 200/1800 & 250/250 & 25/50 \\
KASKA (Japan) & 24 & 350/1300 & 140/600 & 8.5/8.5 \\
Daya Bay (China) & 11 & 300/1500 & 200/1000 & 20/40 \\
& & & & \\ \hline
\end{tabular}
\end{center}
\end{table}
One of the most promising projects is {\em Double-Chooz}, a
experiment to measure $\theta_{13}$ to be constructed at the Chooz
site. The basic design and features of all the projects are
similar to one another and will be illustrated with Double-Chooz.

\section{Double-Chooz: An Example of Generic Design}

\begin{figure}[t]
\begin{center}
\psfig{figure=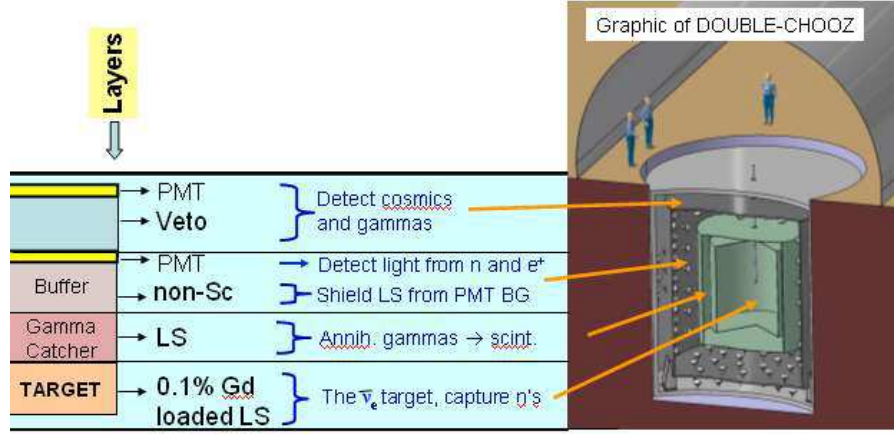,scale=0.6}
\end{center}
    \caption{Artist conception of the {\it Double-Chooz} project showing the layers of the
    various components (left) of the detector~\protect \cite{loi}}
    \label{f:generic}
\end{figure}

The basic structure of {\it Double-Chooz} is illustrated in
Figure~\ref{f:generic}.
The important features of the design are:
\begin{itemize}
    \item Gd loaded liquid scintillator (LS) central target defining the fiducial volume
    \item LS surrounding target to catch annihilation gammas (Gamma Catcher)
    \item Non-scintillating buffer to shield target from
    the photomultipliers (PMT's)
    \item Passive or active veto to detect cosmics
    \item Two detectors to cancel reactor related systematics
    \item Deep underground to decrease cosmogenic backgrounds
\end{itemize}
Details of the {\it Double-Chooz}\ project are described in the
next section~\footnote{Most of the information in the next section
is taken from the Double Chooz Letter of Intent~ \protect
\cite{loi}}.

\subsection{Anti-Neutrino Calorimetry}

\begin{center}
\begin{figure}[t,h,b]
\centering \psfig{figure=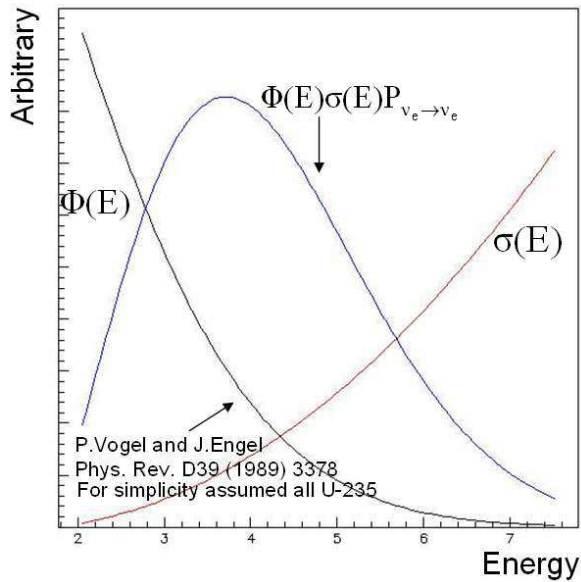,scale=0.4}
    \caption{Figure showing the neutrino flux (calculated for $^{235}$U only)
    and the cross-section for inverse beta decay as a function of
    energy and the positron spectrum.}
\label{f:spectrum}
\end{figure}
\end{center}

Anti-neutrinos are produced via the beta decay of the nuclear fuel
used in the reactors. The antineutrino can react with a free
proton via inverse beta decay:
\begin{eqnarray*}
\overline{\nu_{e}}\, + p\, \longrightarrow e^{+}\, +\, n
\end{eqnarray*}
The anti-neutrino energy threshold for the inverse beta reaction
is 1.8 MeV. At this energy the positron annihilates at rest and
produces 2 back-to-back gammas. The neutrino flux coming from the
nuclear reactor is an exponentially decaying function of neutrino
energy. Meanwhile, the cross-section of inverse-beta decay is an
exponentially increasing function of energy so that the detected
flux is a spectra peaking at approximately 3.5-4 MeV (see Figure
\ref{f:spectrum}).

\begin{center}
\begin{figure}[t,h,b]
\centering \psfig{figure=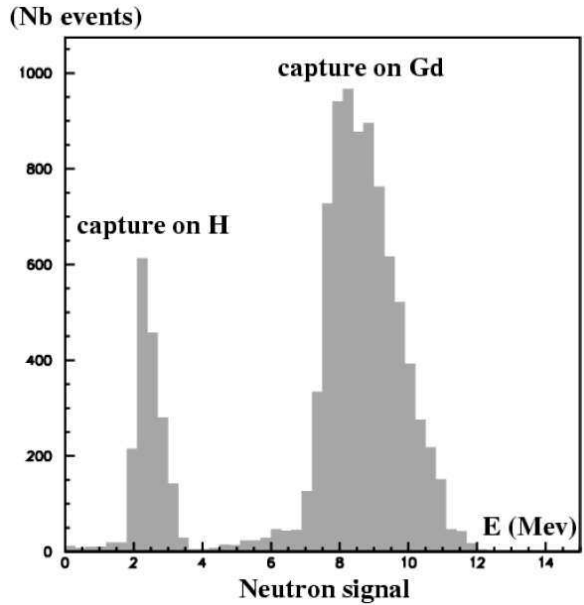,scale=0.4}
    \caption{The gamma spectrum for neutron capture on H and on Gd showing
    the clear separation between the two spectra. Selecting energies exceeding 6 MeV
    will ensure that the neutrino target volume is well defined~ \protect \cite{loi}.}
    \label{f:neutrons}
\end{figure}
\end{center}

In all of the new generation nuclear reactor projects the main
target for the anti-neutrinos is a Gd doped LS volume (see
Figure~\ref{f:generic}). Natural Gd has a very high neutron
capture cross-section (49000 barn averaged over the
isotopes~\cite{karls}) and produces gammas peaking at $\sim$8 MeV
when the excited states decay. Therefore, a specific signature for
the detection of an electron antineutrino is the detection of
prompt gamma's from the annihilation of the positron and the
delayed capture of a neutron several tens of $\mu$s later. The
gammas are detected by the central LS target, and the surrounding
LS known as the 'Gamma-Catcher'. It should be mentioned that the
neutrons produced in the target may also be captured in the free-H
present in the LS in both the target and the Gamma-catcher, and
this would create uncertainty in determining the neutrino target
volume. However, the neutron capture gamma spectrum for H is very
different from Gd in that the H high energy tail does not exceed
$\sim$4 MeV. Therefore, the target fiducial volume can be selected
by accepting events of energy greater than 6 MeV. This is
demonstrated in Figure~\ref{f:neutrons}, where the neutron capture
gamma spectrum is shown for {\it Double-Chooz} and the separation
between the H and Gd spectra is evident.

The presence of the Gamma-Catcher ensures that the positron energy
is fully contained, because all the light from the positron
annihilation is detected. An improvement over previous experiments
is the inclusion of a non-scintillating buffer to shield the gamma
catcher and central target from the light produced by the PMT
beta-gammas.  Surrounding the buffer is a cylinder of $\sim$800
PMT's that detect the light from the positron annihilation and Gd
gammas.

\subsection{Two Identical Detectors for Uncertainty Cancellation}

Antineutrinos are produced by nuclear reactors due to the beta
decay of the fission daughters of 4 main isotopes
($^{235}$U,$^{238}$U, $^{239}$Pu, and $^{241}$Pu). In actual fact
there are hundreds of different beta decay spectra to account for
in modelling the total anti-neutrino flux, although the agreement
between theory and experiment is approximately 4\%, it is not
precise enough for the new generation of short baseline nuclear
reactor experiments. Thus, one of the fundamental limitations of
previous experiments is the knowledge of the nuclear reactor
neutrino flux. To improve on this systematic, the new generation
of short baseline reactor neutrino experiments will use two
detectors~\cite{mikaelyan}, one at approximately 100-200 m (Near
Detector) and another detector at $\sim$1-1.2 km tuned to be at
the first $\theta_{13}$ oscillation maximum. The near detector
measures the un-oscillated neutrino spectrum and flux. Therefore,
the nuclear reactor related uncertainties such as neutrino flux,
structure of the core, will not be a limiting factor. The target
volume for the near and far detector for {\it Double-Chooz}\ is
planed to be 12.7 m$^{3}$ and is expected to give a statistical
error $<$0.4\% for the data-set of interest. Furthermore, the
detectors will be identical so that many detector related
efficiencies should cancel. Uncertainties such as scintillator
density, fraction of free protons and hydrogen atom density will
all cancel, since it is planned to fill both the near and far
detector at the same time from the same batch of liquid
scintillator. The important quantity is the relative systematic
error between the near and far detector. This systematic has been
estimated to be $\sim$0.6\%.

\subsection{Further Decrease of Systematics}
The new generation of nuclear reactor neutrino experiments, such
as {\it Double-Chooz} require an extremely small uncertainty to be
sensitive to the subdominant \tai oscillations. This requires not
only cancellation of the detector and reactor related
uncertainties, but also a carefully performed analysis not to
introduce efficiencies and unnecessary cuts, since these all have
accompanying uncertainties. In the Chooz experiment many of the
dominant uncertainties were due to cuts applied when events were
reconstructed. For {\it Double-Chooz} it has been decided not to
apply any reconstruction cuts and this has decreased the number of
cuts from 7 to 3. The remaining systematics related to the cuts
are shown in Table~\ref{t:cuts}.
\begin{table}[t]
\caption{Uncertainties in the analysis cuts and efficiencies for
the {\it Double-Chooz} experiment~ \protect \cite{loi}.}
\label{t:cuts} \vspace{0.4cm}
\begin{center}
\begin{tabular}{|c|c|c|}
\hline
& Chooz & {\it Double-Chooz}  \\
\hline & & \\
Proton Density & 0.8 \% & 0.2 \% \\
Neutron Efficiency & 0.85 \% & 0.2 \% \\
Neutron Energy Cut & 0.4 \% & 0.2 \% \\
& & \\ \hline
\end{tabular}
\end{center}
\end{table}
When all the analysis cuts and uncertainties are taken into
account the final uncertainty is expected to be
$\sim\pm$0.4(stat)$\pm$0.6(syst)\%.

\subsection{Background Reduction}

The primary feature that distinguishes proposed projects around
the world is the size of the fiducial volume. Groups propose
anti-neutrino targets ranging from 10-50 m$^{3}$, determining a
limit of $\sin^{2}\theta_{13}<0.01-0.03$. However, the size of the
fiducial target is not the only significant feature that
determines the final limit, since high statistics in the far
detector is only useful if the background is low. Anti-neutrino
detection only requires two logical requirements:
\begin{itemize}
    \item Condition 1: Two pulses within 200 $\mu$s
    \item Condition 2: Second pulse must have E$>$6 MeV
\end{itemize}
Therefore, the backgrounds to this signal are accidental and
correlated coincidences produced by natural radioactivity and
cosmic ray induced events. The effect of cosmic rays can be
sufficiently reduced by constructing the detectors deep
underground to reduce the cosmic ray flux. In addition, in the \dc
design there is a muon veto anti-coincidence shield. The muon veto
consists of a LS buffer surrounded and optically separated from
the PMT support structure. Surrounding the buffer there are
additional PMTs which point inward so that cosmic ray muons can be
tagged when these traverse the buffer. Therefore, by careful
material selection, placing the detectors underground, and
implementing an efficient muon veto the goal of \dc is to reduce
the background to 1\% of the signal. Some of the components of the
background are:
\begin{itemize}
    \item {\it Accidental background produced by $^{232}$Th, $^{238}$U,
$^{40}$K present in the detector materials.} The concentration of
Th/U in the scintillator and acrylic vessel must be less than
10$^{-12}$ and 10$^{-10}$ respectively to ensure this background
is negligible.
   \item {\it Correlated events due to cosmic ray spallation on $^{12}$C in the LS}.
Cosmic ray spallation on the LS can produce various short-lived
isotopes such as $^{9}$Li, $^{11}$Li, and $^{8}$He. These isotopes
subsequently beta decay and can produce neutrons. In some of these
cases the two pulses will have the same time structure as an
anti-neutrino event. The cross-section of $^{8}$He+$^{9}$Li has
been measured by the NA54~\cite{na54} experiment at CERN.
    \item {\it Correlated background due to cosmic ray spallation outside the
    detector.}
In this case the spallation events produce fast neutrons and the
initial muon is not detected by the veto. These neutrons travel to
the central target, and scatter on nuclei to produce fast protons
which in turn generate scintillation light and look like
positrons. The spallation neutron may then be captured in the
target. The timing of such an event looks identical to
anti-neutrino capture. The background for these events can be
calculated using Monte Carlo methods and are expected to be
$<$1/day.
\end{itemize}
The near detector is constructed with an artificial overburden of
60-80 m.w.e. so that the cosmic ray induced backgrounds are much
higher. Fortunately, the increase in background is compensated by
an increase in the signal due to the proximity of the near
detector, so that the background is still $<$ 1\% of the signal.
Furthermore, nuclear reactors tend to be cycled 'on' and 'off'
when maintenance is done, or when fuel rods are changed, so that
this will allow a 'beam-off' measurement to be done. In this way
the background can be independently measured.

\subsection{$\theta_{13}$ Sensitivity}

\begin{center}
\begin{figure}[t,h,b]
\centering \psfig{figure=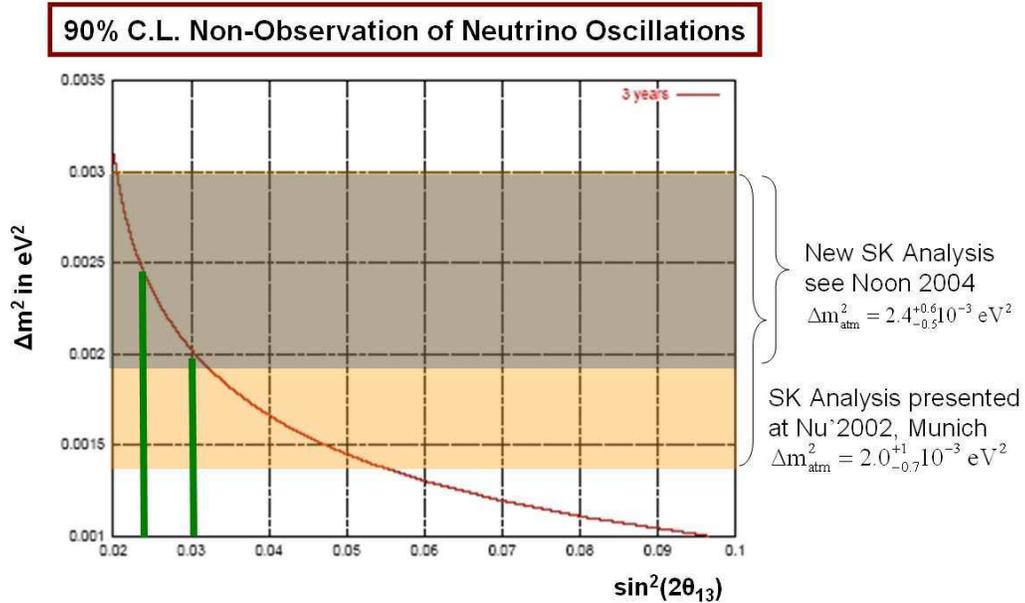,scale=0.7}

\caption[]{The 90 \% C.L. upper limit that can be placed on \ssq
as a function of \dmatm for a 3 year data set\, \protect
\cite{loi}. } \label{f:sensitivity}
\end{figure}
\end{center}

The {\it Double-Chooz} sensitivity is calculated by doing a
$\chi^{2}$ fit of the expected number of events for various
oscillation scenarios compared to the no-oscillation case. This
analysis takes into account the relative normalization error,
energy and spectral shape uncertainty, fluctuation in the core and
background subtraction error. The \ssq limit for no-oscillations
is a strong function of \dmatm\ in the sense that if \dmatm\ is
higher but no events are observed, the \ssq limit is pushed lower.
Recently SK presented a new analysis to estimate \dmatm\ by
reconstructing the energy and path length (using direction with
respect to zenith angle) of atmospheric neutrinos. They proved
from the (L/E) analysis that the data follows the 'dip' expected
due to \tsol\ oscillations and extracted
\dmatm=2.4$^{+0.6}_{-0.5}\times$10$^{-3}$ eV$^{2}$, an increase
from the value presented at Neutrino 2002 in Munich. The no
oscillation sensitivity for {\it Double-Chooz} is presented in
Figure~\ref{f:sensitivity} as a function of \dmatm, the shaded
regions shows the two SK analyses. The sensitivity of the various
other proposed reactor experiments to \tai\ ranges from
\ssq$<$0.01-0.03. In the case of discovery, a non-vanishing
\ssq=0.05(0.04) at 3$\sigma$ could be measured in 3 years of data
running for \dmatm=2.0$\times$10$^{-3}$
eV$^{2}($2.4$\times$10$^{-3}$ eV$^{2}$).

\section{Conclusions}

The past few years have seen a lot of progress in a new generation
of nuclear reactor experiments specifically designed to measure
$\theta_{13}$. The design requires careful control of systematics
by requiring two detectors, a large overburden, and a baseline for
the far detector tuned to be at the first maximum of the
sub-dominant $\theta_{13}$ oscillations. This allows us to neglect
matter oscillations and make a pure measurement of \ssq. Several
sites around the world are being investigated (see
Table~\ref{t:sites}). Among the most promising is the {\it
Double-Chooz} project, a project planned to be constructed at the
old Chooz detector site. {\it Double-Chooz} plans to set a limit
of \ssq$<$0.03 after 3 years of running.

\end{document}